\documentclass[11pt]{article}

\usepackage{amsmath,amsfonts,amssymb,amsthm,amsxtra,bbm,upref,graphicx,epsfig,mathrsfs,tabularx,bm,times,color, multirow}
\usepackage{verbatim}
\usepackage{booktabs}
\usepackage[numbers]{natbib}
\usepackage{tikz}
\usetikzlibrary{arrows.meta}
\tikzset{%
	>={Latex[width=2mm,length=2mm]},
    pil/.style={
          ->,
          thick,
          shorten <=2pt,
          shorten >=2pt,},
	base/.style = {rectangle, rounded corners, draw=black!0,
		minimum width=4cm, minimum height=1cm,
		text centered, font=\sffamily},
	base1/.style = {rectangle, rounded corners, draw=black,
		minimum width=4cm, minimum height=1cm,
		text centered, font=\sffamily},
	graybox/.style = {fill=black!10, draw=black!10},        
	bluebox/.style = {base1, fill=blue!20},
	redbox/.style = {base1, fill=red!20},
	greenbox/.style = {base1, fill=green!30},
	orangebox/.style = {base1, minimum width=2.5cm, fill=orange!15,
		font=\sffamily},
}
\usepackage{xcolor}
\usepackage{graphicx}
\usepackage{url}
\usepackage{verbatim}
\usepackage{subfigure}
\usepackage{rotating} 
\usepackage{csquotes} 
\usepackage{epigraph}
\usepackage{recycle} 
\usepackage{soul} 

\newcommand{\be}{\begin{equation}}
	\newcommand{\enq}{\end{equation}}

\makeatletter
\newcommand*{\indep}{%
	\mathbin{%
		\mathpalette{\@indep}{}%
	}%
}

\usepackage{epstopdf}
\AppendGraphicsExtensions{.tif}

\begin{document}

\title{Extent of Safety Database in Pediatric Drug Development: Types of Assessment, Analytical Precision, and Pathway for Extrapolation through On-Target Effects}

\author{Margaret Gamalo\footnote{$^{\dagger}$Corresponding Author: Margaret Gamalo, Pfizer, Collegeville, PA; Email:margaret.gamalo@pfizer.com.com. The views expressed in this paper are those of the authors and not necessarily those of the author's employers.}, Yihua Zhao, Aijun Gao, Jingjing Ye, Ralph DeMasi, \\ 
Eiji Eshida, YJ Choi, Robert Nelson
}
\date{}

\maketitle

\begin{abstract}
Pediatric patients should have access to medicines that have been appropriately evaluated for safety and efficacy. Given this goal of revised labelling, the adequacy of the pediatric clinical development plan and resulting safety database must inform a favorable benefit-risk assessment for the intended use of the medicinal product. While extrapolation from adults can be used to support efficacy of drugs in children, there may be a reluctance to use the same approach in safety assessments, wiping out potential gains in trial efficiency through a reduction of sample size. To address this reluctance, we explore safety review in pediatric trials, including factors affecting these data, specific types of safety assessments, and precision on the estimation of event rates for specific adverse events (AEs) that can be achieved.  In addition, we discuss the assessments which can provide a benchmark for the use of extrapolation of safety that focuses on on-target effects. Finally, we explore a unified approach for understanding precision using Bayesian approaches as the most appropriate methodology to describe/ascertain risk in probabilistic terms for the estimate of the event rate of specific AEs. 

\end{abstract}


\section{Introduction}
Pediatric patients should have access to medicines that have been appropriately evaluated for them. Towards this goal, legislations have been enacted in several jurisdictions to encourage, incentivize, and require studies of investigational drugs in children when pediatric use is anticipated \citep{bucci2017enhancing}. As a result, pediatric research has been stimulated across the globe in the last three decades, with FDA (Food and Drug Administration) making labeling changes in 854 drugs as of April 2020 \citep{FDA_peds_label}. To reduce the exposure of children to unnecessary clinical trials, both the U.S (United States). and EU (European Union) regulations suggest an extrapolation strategy based on assessing the relevance of existing information in the adult population to the pediatric population, in terms of the similarity of disease, drug pharmacology and clinical response to treatment to identify the gaps or level of uncertainty that need to be addressed to extend from adult to pediatric patients the conclusion of adequate evidence of efficacy and safety \citep{european2012reflection, ICHE11R1}. This strategy is called {\it pediatric extrapolation} and is adopted broadly by ICH (International Council for Harmonization) member parties to maximize the use of pre-existing information and to reduce the amount of, or general need for, additional information to reach conclusions of effective and safe use of drugs in children \citep{ICHE11A}. The recent ICH E11A draft guidance on the use of pediatric extrapolation explicitly discusses how this concept can be extended to the assessment of pediatric-specific safety for establishing positive benefit-risk of the drug in children. 

As with any drug development program, the U.S. evidentiary standard (21 CFR 314.50) for approval of a medicinal product for pediatric patients is the same as in adults, i.e., the product needs to be safe and effective for labeled indications under labeled conditions of use. Identifying and assessing the safety reported in pediatric clinical trials is important in determining whether a product’s labeling needs to be revised \citep{rodriguez2008improving}. In drug reviews, reviewers may explicitly note whether the observed treatment-related adverse events in children are reflected in the existing labeling (for previously approved products), or whether some revisions are needed. This observation is of particular importance because the disease pathophysiology, along with growth and developmental changes, can create unanticipated dose-response, adverse events (AE), or clinically significant laboratory changes in the pediatric compared to the adult patient \citep{mcmahon2018assessing, kim2021drug}. It also calls for data needed to support such conclusions.  

While pediatric trial sample size usually is driven by the objective of establishing efficacy, there is no clear benchmark to guide the sample size necessary for assessing safety. Unlike efficacy where a clear primary endpoint can be named, the evaluation of safety is multi-objective where the common assessments are treatment emergent AEs (TEAEs), serious AEs (SAEs), severe AEs, and AEs leading to discontinuation of the participant from study. These adverse events may also be associated with many quantities or qualities of interest, e.g., incidence, timing, duration, reversibility, recurrence of AEs, the impact of dose (reduction), timing and frequency of treatment, effects of rescue treatments, among others. In addition, while the ICH E1A has made recommendations for adequate safety database in adults for drugs taken chronically, some of the recommendations may not be feasible in children (see \citep{ICHE1A}). Because of the limited trial size in most pediatric clinical trials, there is a need to prioritize what objectives can be de-risked while keeping track of other risks and their attendant precision. In relation to the goal of labelling revision, it may be important to look at two areas- on-target effects and pediatric developmental or growth and maturation domains. It is also important to ascertain how that precision can be increased or supplemented in post-marketing activities. 

Given that adverse events are infrequent in children, the determination of safety may require a large extent of development and will make trials infeasible. Of note, even in the post-marketing setting, \citet{hwang2019completion} reported that only 33.8\% of pediatric studies mandated by the Pediatric Research Equity Act have been completed after a median follow-up of 6.8 years, and 41.2\% of drug labels include any pediatric data. The recently drafted ICH E11A discussed extrapolation of safety data which can be considered based on the available knowledge of the known and/or potential safety issues in the reference population that are relevant to the target pediatric population. Other information (e.g. nonclinical, mechanistic) should be considered as part of this analysis \citep{ICHE11A}. These focused extrapolation of safety data can help increase certainty about the expected safety profile of a drug in a particular pediatric population and provide guidance if additional gaps in knowledge need to be addressed in the pediatric program.

In this paper, we explore the extent of the safety database in pediatric trials. We explore factors affecting this database, specific safety assessments in children, and what assessments may be useful for driving the extent of development. We explore precision, the proper hypothesis that needs to be tested, and how large the safety databases need to be for the conclusion to be reliable. We explore the use of Bayesian approaches as the most appropriate methodology to describe/ascertain risk in probabilistic terms with available data. We provide unified expressions for the assessment of key quantities of interest such as confidence level, incidence rate, and observed data. Finally, we explore the potential for the use of extrapolation for specific types of adverse events, e.g., on-target effects, where these are expected in any population.  

\section{Specific types of pediatric safety assessments}
Adverse events may be classified as belonging to a certain type according to their reaction profiles: augmentation of normal drug effects (Type A), bizarre effects (Type B), chronic effects (Type C), delayed effects (Type D), and end of drug use effects (Type E) \citep{elzagallaai2017adverse}. Augmentation of normal drug effects is often characterized by exaggerated and adverse pharmacologic effects when a drug interacts with its intended receptor at the intended tissue of interest (``on-target effects''). Often on-target effects are identified at compound target validation which includes various queries to identify and define potential roles that the target could play in normal physiologic processes or homeostasis leading to early hypothesis testing for target-related safety risks (\citep{rudmann2013target}). It is through on-target effects that it is determined whether the potential on-target safety liability fits the indication (benefit/risk), i.e., right safety for the right patient \citep{morgan2018impact}. On-target effects are predictable, tend to be mild, can resolve with dose reduction, and constitutes about 80\% of all observed AEs \citep{elzagallaai2017adverse}. On-target effects may also occur when a drug interacts with its intended receptor, but in an unintended tissue \citep{golan2011principles}. Most on-target effects are considered as adverse drug reactions (ADRs) an undesirable effect, reasonably associated with the use of a drug, that may occur as part of the pharmacological action of the drug or may be unpredictable in its occurrence \citep{FDA_safety_review}. If these on-target AEs are subject to extrapolation, it is of interest to anchor the extent of the safety database on the predictable drug reactions that can be anticipated when planning a pediatric development program and whether the extent of safety database can provide information of an increase in these ADRs in children which impacts benefit-risk determination. This approach can help increase certainty about the expected safety profile of a drug in a particular pediatric population.

The ability to estimate and provide adequate precision for Type B or idiosyncratic/bizarre adverse events is challenging as they often occur less frequently. The number of these so-called ``off-target'' effects for a small molecule is probably always significantly higher than on-target effects because profiling of compounds uses a predefined (restricted) non-overlapping target panel consisting of a limited set of targets (typically $<100$) \citep{lynch2017potential}. In addition, compounds that bind to numerous, unintended targets are associated with greater toxicity. While off-target interactions are generally weaker in affinity than those with the intended pharmacological target, they may be relevant in cases of higher cellular expression of the off-target or high systemic exposure, such as in preclinical toxicity studies (where higher doses are interrogated to define the toxic profile of the compound), clinical mis-dosings, accidental or intentional overdose, drug-drug interactions (which may lead to higher systemic exposures), or other unanticipated individual variations (which also can lead to higher systemic exposures) \citep{whitebread2016secondary}. These events are often unpredictable, have no clear dose-dependency, and account for about less than 20\% of all adverse events observed \citep{elzagallaai2017adverse}. 

\begin{figure}[ht!]
     \begin{center}
            \includegraphics[width=4in]{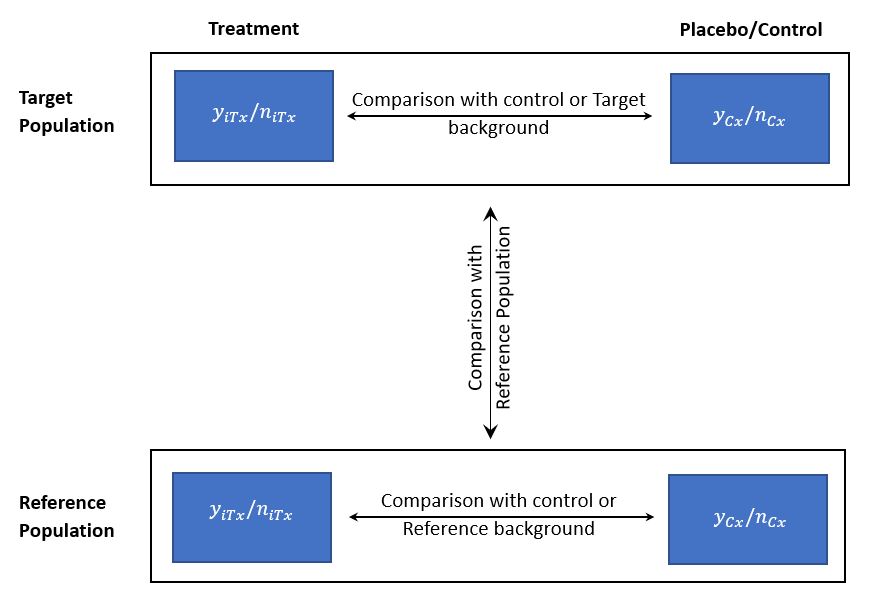}
    \end{center}
    \caption{Typical comparison in safety assessment for adverse drug reactions in children
     }%
   \label{fig:Safety_Comparison}
\end{figure}

We do not always fully understand the extent of the mechanism of toxicity and safety; however, a battery of in vitro pharmacology assays is usually tested to determine off-target interactions \citep{lynch2017potential}. It is proposed that the extent of safety database can be driven by the ability to characterize known on-target effects in children and determine whether there is high likelihood that they are similar to adults. In particular, the rates in children should not negatively change the benefit-risk calculus in the intended population, e.g., an increase of more than 3 times the rate in adults (see Figure \ref{fig:Safety_Comparison}). This is directionally aligned with how drug candidates are selected using the principle of the right target for the right patients \citep{morgan2018impact}. The use of on-target effects also makes it convenient to improve the precision of estimates since there is a possibility to leverage other related populations. Furthermore, the assessment of these event rates generally comes from well-controlled trials enabling causal estimation of effects. The proposal does not mean, however, that the adverse drug reaction is the only assessment that will be conducted and that assessments for other types of AEs must still be part of the safety assessment routine. As a rule, and like the safety assessments conducted in the adult population, the strategy in children should still be find ways to minimize or find mitigation for the pharmacological engagement of unintended molecular target(s); and/or other non-pharmacological toxicity (e.g. membrane damage). 

Alternatively, but of equal importance, the extent of the safety database may be related to ensuring that safety risks related to pediatric development are rigorously assessed. Depending on the age of the target population and the mechanism of action of the drug, data on developmental parameters, e.g., growth, reproductive and neurocognitive development need to be collected. Abnormal findings in juvenile toxicology studies can guide in ensuring proper assessments are planned and captured in clinical trials. In terms of analysis, if the disease does not impact development, one way to evaluate a potential effect of drug exposure on growth may be by individually comparing study participants in weight measurements at each monthly age over time against the general reference pediatric population, which information is available at CDC (Centers for Disease Control) Growth Charts and WHO (World Health Organization) Growth Standards. By following up individual time course of body weight with respect to the Growth Chart percentiles at each monthly age time point, we may identify an effect of a new treatment on body weight and its change over time. Such time profiles of body weight can help the investigators identify a potential growth issue of each patient in the use of a new treatment. Such profiles may also make a basis for an analysis of the drug safety effect in the study population when they are combined for each monthly age, for instance, although individual growth status may differ at each monthly age (see recommendations in \citep{choi2022}). 

\section{Extent of safety database through precision in assessments types}

To explore extent of database for given precision, suppose $y_{gz}$ as the number of events out of $n_{gz}$ patients in the safety analysis set for each arm $z\in\{C_x,\ iT_x\ \}$, $\mathcal{G}=\{1,\ldots,\ G\}$. Let $\theta_{g,z,k}$ be the parameter of interest corresponding to the incidence proportion of the specific AE indexed by $k$ in the treatment arm $z$ for cohort $g\in \mathcal{G}$. While alternative measures can be used, e.g., an incidence rate that accounts for how long each patient is exposed, the exposure times of patients between the investigational treatment arm and placebo are similar in a placebo-controlled period of the trial. Hence, exposure adjustment may not be necessary to obtain a good estimate of the placebo-corrected incidence rates. 

\subsection{Precision for Common Identified/Potential Risk Regardless of Importance}\label{ADR}
The basis for inclusion in the Adverse Reaction section of the United States Package Insert (USPI) is that the AE occurs $> 1\%$ in the treated group and for which the rate for a drug exceeds the rate for placebo \citep{FDA_safety_review}). Adverse drug reactions are generally known/identified or potentially related to the mechanism of action and they are continually monitored in the development. An identified risk is an untoward occurrence for which there is adequate evidence of an association with the medicinal product of interest \citep{FDA_safety_review}. Many ADRs are also on-target effects and in reviews of marketing authorizations for a drug in a pediatric indication, there generally is a mention of a comparison of ADRs between adults and target pediatric cohort. Others note differences and sometimes do not mention of labeling changes at all. In these assessments, the hypothesis that is of interest is of similarity or that there is no clinically relevant difference in the incidence of a specific ADR or on-target effect in children. Suppose $g\in \{1,2\}$, where $g=1$ refers to the reference population and $g=2$ is the target population. Then consistency of the incidence rates or an ``equivalence testing'' is more suitable, i.e., 
\begin{equation} \label{eq:equivalence}
H_{0,k}: \vartheta_{2,k} \geq \hat{\vartheta}_{1,k}-\varepsilon \quad  vs  \quad H_{1,k}:\vartheta_{2,k} < \hat{\vartheta}_{1,k}+\varepsilon,
\end{equation} where $\vartheta_{2,k}=\theta_{2, iTx,k}-\theta_{2, Cx,k}$, for $0 <\varepsilon \leq \hat{\vartheta}_{1,k}-z_{\alpha}\mathrm{var}(\vartheta_{1,k})$, and $k$ indexes the ADRs. In other instances, consistency is defined as 
\begin{equation} \label{eq:multiples}
H_{0,k}: \vartheta_{2,k} \leq f\hat{\vartheta}_{1,k} \quad vs \quad H_{0,k}: \vartheta_{2,k} > f\hat{\vartheta}_{1,k}
\end{equation} where $f>1$ corresponding to a multiple of the rate observed in the reference population. Of note, because $\vartheta_{2,k}$ is small, the second hypothesis (\ref{eq:multiples}) may be more appropriate as it may be challenging to establish a margin about a small rate. 

For ease of exposition and notation, we will drop the index $k$. If the initial prior for $\theta_{2z}$, for the risk difference $\vartheta_{2}=\theta_{2, iTx}-\theta_{2, Cx}$, is $\pi_{\theta_{2z}}=\ Beta(a_{z\prime},\ b_{z\prime })$ then the posterior distribution for $\theta_{2z}$ given observed data $(r_{2z}, n_{2z})$ is $q_{\theta_{2z}}=Beta(r_{2z}+a_{z\prime },n_{2z}-r_{2zk}+b_{z\prime })$. This observed data could be any available information on the ADR before conducting the pediatric trial. Let $a_{z}= r_{2z}+a_{z\prime}$, $b_{z}= n_{2z}-r_{2z}+b_{z\prime }$ so that the distribution $p_{\vartheta_2}(x)$ can be expressed using parameters $a_{z},\ b_{z}$ which if $z\in \{iTx\equiv 1, Cx\equiv 2\}$ has the form   
\begin{equation}
    p_{\vartheta_2}(x) = 
  \begin{cases}
            A^{-1}B(a_2,\ b_1)\ x^{b_1+b_2-1}\ (1-x)^{a_2+b_1-1} &\\
            \quad\quad \times F_1(b_1,\ a_1+a_2+b_1+b_2-2,\ 1-a_1;b_1+a_2; & \\
            \quad\quad\quad 1-x, 1-x^2)  & \text{if $0\le x<1$} \\
            A^{-1}B(a_1+a_2-1,\ b_1+b_2-1) & \text{if $x=0,\ a_1+a_2>1,\ b_1+\ b_2>1$} \\
  A^{-1}B(a_1,b_2)\ (-x)^{b_1+b_2-1}(1+x)^{a_1+b_2-1} & \\
  \quad \quad \times F_1(b_2,\ 1-a_2, a_1+a_2+b_1+b_2-2;a_1+b_2;& \\
  \quad\quad\quad 1-x^2,1+x) & \text{if $-1<x\leq 0$}
  \end{cases}
\end{equation} defined on the interval (-1,1) (see \citep{pham1993bayesian}). Here $A=B(a_1,\ b_1)B(a_2,\ b_2)$ and $F_1\ (\cdot)$ is Appell’s first hypergeometric function computed as 
\begin{equation}
F_1(u,v_1,v_2;w;x_1,x_2)=\sum_{i=0}^\infty\sum_{j=0}^\infty \frac{(u)^{i+j}(v_1)^{i}(v_2)^{j}}{(w)^{i+j}} \frac{x_1^ix_2^j}{i!j!}
\end{equation} 
with Pochhammer symbol $(r)^s=\Gamma(r+s)/\Gamma(r)$ (see \citep{pham1998distribution}). The double series representation is convenient for finding the confidence level or probability $C\equiv Pr(\vartheta_2<f\hat{\vartheta}_1)=\int_{-1}^{f\hat{\vartheta}_1} p_{\vartheta_2}(x)dx$, for the estimate of the specific adverse event in children not exceeding a certain multiple of what $\hat{\vartheta}_1$ was observed in adults. This probability can be split into the following depending on where $f\hat{\vartheta}_1$ is in the $[-1,\ 1]$ interval, i.e., 
\begin{eqnarray}\label{eq:confidence1}
C\equiv Pr(\vartheta_2<f\hat{\vartheta}_1)& =& \int_{-1}^{f\hat{\vartheta}_1}A^{-1}B(a_2,\ b_1\ )\ x^{b_1+b_2-1} (1-x)^{a_2+b_1-1}\nonumber \\
& & \times F_1\ (b_1,\ a_1+a_2+b_1+b_2-2,\ 1-a_1;b_1+a_2;\ 1-x,\ 1-x^2) dx 
\end{eqnarray} for $0\leq f\hat{\vartheta}_1<1$
and 
\begin{eqnarray}\label{eq:confidence2}
C\equiv Pr(\vartheta_2<f\hat{\vartheta}_1)& =& 1-\int_{f\hat{\vartheta}_1}^{1}A^{-1}B(a_1,\ b_2\ )\ (-x)^{b_1+b_2-1}(1+x)^{a_1+b_2-1}\nonumber \\
& & \times F_1(b_2,\ 1-a_2, a_1+a_2+b_1+b_2-2;a_1+b_2; 1-x^2, 1+x) dx
\end{eqnarray} for $-1\leq f\hat{\vartheta}_1<0$. The relationship of the quantities of interest $C$, $\hat{\vartheta}_1$, $f$, $(a_z, b_z)$ can be characterized from equations (\ref{eq:confidence1})-(\ref{eq:confidence2}). These integrals can then be approximated through the trapezoidal rule or Monte Carlo methods to avoid numerical tolerance level issues. Alternatively, asymptotic normal approximations on each of the posterior distributions of $\vartheta_{2,z}$ can also be used if there is difficulty in computing the above expressions. A very low probability would indicate that the estimates of the placebo corrected event rate for a specific adverse event in a pediatric cohort is not consistent with what was observed in its respective reference population. Of note, the case of single arm trials is discussed in the subsequent section. These probabilistic assessments provide information on the degree of current uncertainty and how information and extent of post-marketing surveillance can be guided appropriately to increase precision of estimates based on current information.

\begin{figure}[ht!]
     \begin{center}
        \subfigure[Background Pbo inc. prop = 0.003]{%
            \includegraphics[width=3.1in]{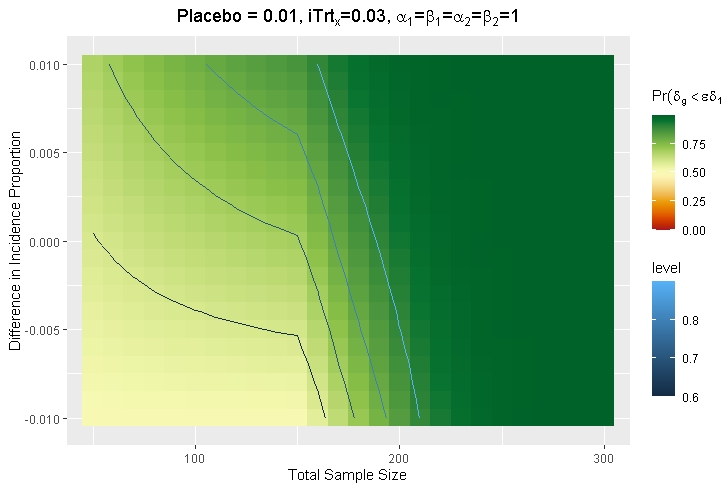}
        }%
        \subfigure[Background Pbo inc. prop = 0.01]{%
            \includegraphics[width=3.1in]{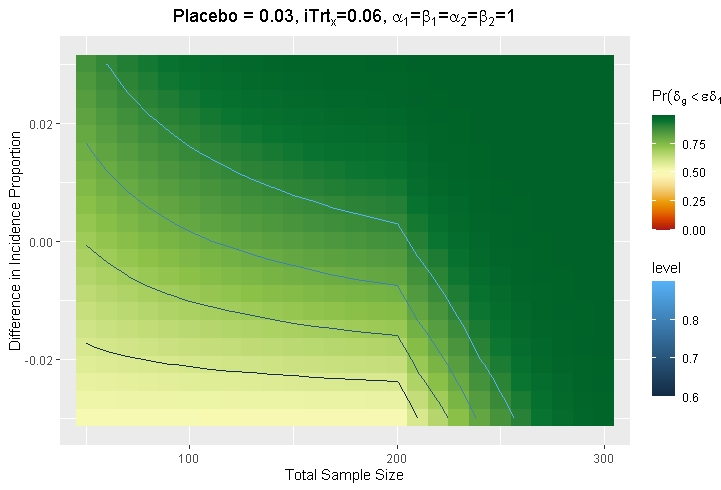}
        }\\
    \end{center}
    \caption{ Probability of consistent placebo corrected incidence proportion in the target children's cohort relative to reference cohort. }%
   \label{fig:prob_equivalence}
\end{figure}

\begin{figure}[ht!]
     \begin{center}
        \subfigure[Background Pbo inc. prop = 0.0025]{%
            \includegraphics[width=3in]{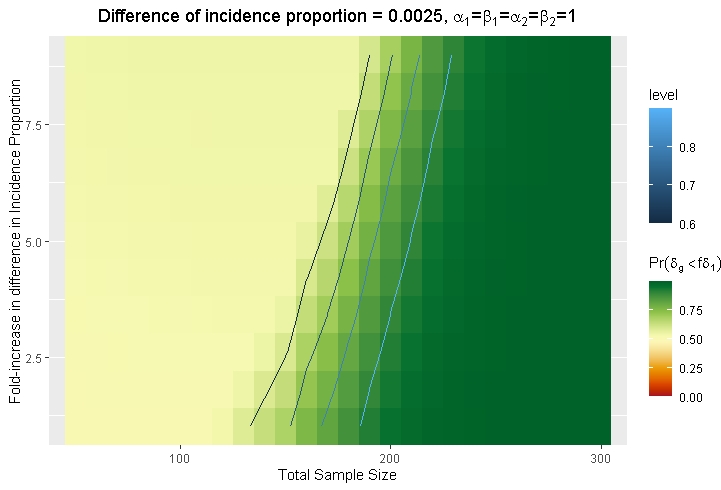}
        }%
        \subfigure[Background Pbo inc. prop = 0.01]{%
            \includegraphics[width=3in]{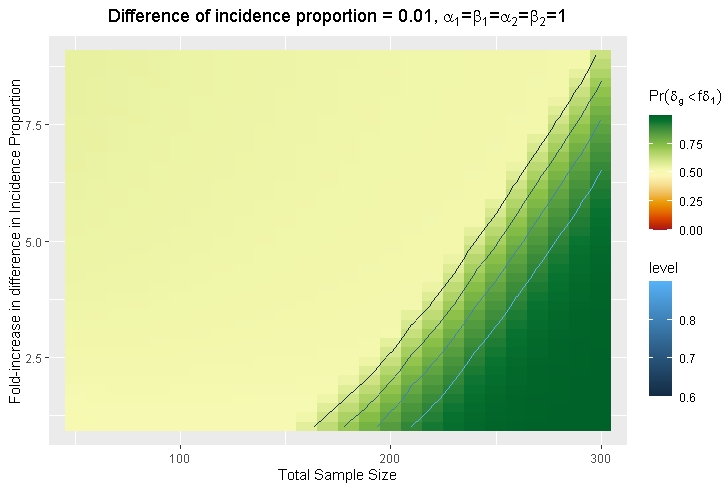}
        }\\
        \subfigure[Background Pbo inc. prop = 0.005]{%
            \includegraphics[width=3in]{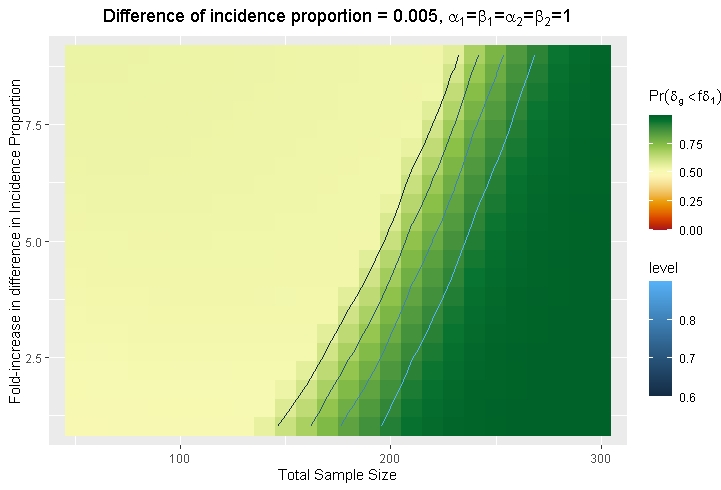}
        }%
        \subfigure[Background Pbo inc. prop = 0.03]{%
            \includegraphics[width=3in]{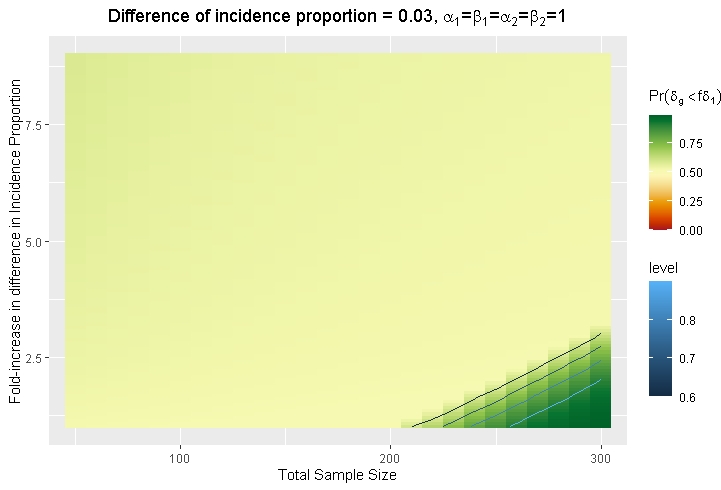}
        }\\
    \end{center}
    \caption{ Probability of ruling out an $f$-fold increase in the difference in incidence proportion relative to placebo.}%
   \label{fig:prob_ruleout}
\end{figure}

Figure \ref{fig:prob_equivalence} shows the chance of ruling out clinically meaningful increase ($\varepsilon$) in the placebo corrected AE rate at different common background rate according to hypothesis (\ref{eq:equivalence}). For example, suppose for example that a specific common ADR was observed to have a difference in incidence proportion of 3\% in the pediatric population given the investigational drug when the background rate is 1\%, then to rule out that there is an $\varepsilon = 0.5\%$ increase in the placebo corrected rate with a probability of 80\%, a sample size of approximately 180 pediatric participants randomized 1:1 is needed between the treatment and placebo control. As seen in Figure \ref{fig:prob_equivalence}, when the extent of the safety database is small, there is a significant probability that a meaningful absolute increase in the incidence proportion cannot be ruled out. In fact, to be certain that there is no increase, the placebo-corrected rate in the pediatric cohort has to be lower than the reference population. Figure \ref{fig:prob_equivalence} also shows a steep increase in information at certain sample size associated with a particular background rate. If the background rate gets higher, this sample size requirement gets larger. Hence, if there is some evidence that a serious ADR has an increased incidence, the probability assessments show that it may be important to ensure that the extent of the safety database is increased proportionally to the level of confidence we can be assured of the precision of the estimate.  

Figure \ref{fig:prob_ruleout} shows the chance of ruling out different multiples at different common background rates per hypothesis (\ref{eq:multiples}). For example, suppose for example that a specific bizarre effect was observed to have a difference in incidence proportion of  1\% in the pediatric population when the background rate is 1\%, then to rule out that there is no doubling of that placebo corrected rate with a probability of 80\%, a sample size of more than 200 pediatric participants randomized 1:1 is needed between the investigational treatment and placebo control. To rule out that there is a tripling of that rate with probability of 80\%, more than 250 participants are needed (see  Figure \ref{fig:prob_ruleout}). If there does not seem to be a difference in adults and children, the rate in children can be made more precise by using the adult information as prior and potentially reducing the extent of the safety database needed.  Of note, if testing is expanded to include any adverse event, any noticeable increase in incidence relative to placebo and over adults will need to be evaluated whether it rises to the level of an important identified risk. Poisson regression can also be used if the exposure time is used for adjustment. 

Since there will be many important identified or potential risks, one may also focus only on the medically most serious, i.e., important identified or potential risks can be ranked according to medical seriousness or ranked according to severity and disease burden (see for e.g. \citep{seifu2022}). These can then be analyzed through prioritized outcomes (see \citep{finkelstein1999combining, pocock2012win}). In particular, \citet{pocock2012win} introduced the concept of the win ratio (WR) in the analysis of composite endpoints based on clinical priorities. Pairs of subjects are compared based on prioritized component outcomes. The win ratio statistic is then calculated as the number of ``winners’’ divided by the number of ``losers’’ in the test group. To make this operational in terms of testing similarity in the context of the assessment of safety, the win odds (WO) can be used instead where ties are handled by counting them as a half win to each subject within a pair \citet{dong2019win}. In particular, if $\Psi$ is the $\mathrm{WO}$ parameter. The NI hypothesis for $\Psi$ is written as $H_0:\Psi\leq\theta$ vs $H_0:\Psi<\theta$, where $\theta$ is NI margin with $0<\theta<1$. For example, when $\theta=0.8$, the null hypothesis is that the probability of winning for a subject in the test group is less than or equal to 80\% of that for a subject in the control group, with ties splitting evenly. While this technique amplifies events to increase statistical power, it will not give probabilistic statement as it does not involve a posterior distribution but a testing procedure whether there is sufficient evidence for the dissimilarity between the rate of the specific adverse event of interest or not.

\subsection{Precision for Uncommon Important Identified/Potential Risks}

Less common important identified/potential risks or adverse events of interest are usually observed in the context of long-term extension trials. In these trials, a placebo may no longer be present as exposing a child to no treatment for a long time may not be ethical or will render the trial not competitive and infeasible. In this case, control is no longer available and analysis usually involves characterizing incidence proportions and exposure time. Poisson regression or Cox regression are used when exposure to the drug is incorporated. In labeling, however, exposure time is not usually included and only crude incidences are stated and so we will focus the discussion on incidence proportions and its associated posterior distribution. The discussion on assessment of precision in rates using Poisson or hazard ratios which incorporates exposures will be a future discussion and beyond the scope of this manuscript. 

Because a small increment in the incidence proportion cannot be associated with deleterious effects absent a control, the hypothesis being tested here is ruling out multiples of incidence proportion in the reference population, i.e., consistency defined as 
\begin{equation} \label{eq: multiples_single} 
H_{0,k}: \theta_{2,k} \leq f\hat{\theta}_{1,k} \quad vs \quad H_{1,k}: \theta_{2,k} > f\hat{\theta}_{1,k}
\end{equation} where $f>1$ corresponding to a multiple of the rate observed in the reference population. In addition, it is also of interest whether there were unique safety risks identified in children and that if none are seen, what is the associated confidence level. Suppose $y_{2}$ as the number of events out of $n_{2}$ patients in the safety analysis set for the investigational treatment arm, $g \in \{1,2\}$. Let $\theta_{2}$ be the parameter of interest corresponding to the incidence proportion of the specific event  for the target population. If the prior for $\theta_{2}$ is $\pi_{\theta_2}=\ Beta(a\prime,\ b\prime)$ then the posterior distribution for $\theta_{2}$ given observed data $(r_{2}, n_{2})$ is $q_{\theta_2}=Beta(r_{2}+a\prime,n_{2}-r_{2}+b\prime)$. Then quantities of interest $C$, $\theta_1$, $f$, and $r_2$ can be completely characterized from 
\begin{equation}\label{eq: single_beta}
C\equiv Pr(\theta_2<f\hat{\theta}_1|n_2,r_2) = \frac{1}{B(r_2+a', n_2-r_2+b')}\int_0^{f\hat{\theta}_1}x^{r_{2}+a'-1}(1-x)^{n_2-r_2+b'-1}dx
\end{equation} which is related to the problem explored by \citet{liu2016simple}. There are suggestions that because these events are infrequent that the use of a ``near-zero informative'' prior
\begin{equation}
    \pi_{\theta_2}=\ Beta(a\equiv p_a^2/(1-p_a^2), 1)
\end{equation} is desireable. 
When $0<p_a^2 <0.5$, the prior distribution has a mode close to zero if is close to zero. The distribution is then weighed more towards zero. When $p_a^2=0.5$, the prior follows a uniform distribution. When $0.5 <p_a^2 < 1$, the prior weighs more towards one. Usually, the value of can be based on external information from a similar population, e.g., $p_a^2\in [0.001, 0.1]$. Of note, this prior can also be applied to the case of common adverse reactions in Section \ref{ADR} for adverse drug reactions.

\begin{figure}[ht!]
     \begin{center}
         \subfigure[Probability threshold = 0.70]{%
            \includegraphics[width=3in]{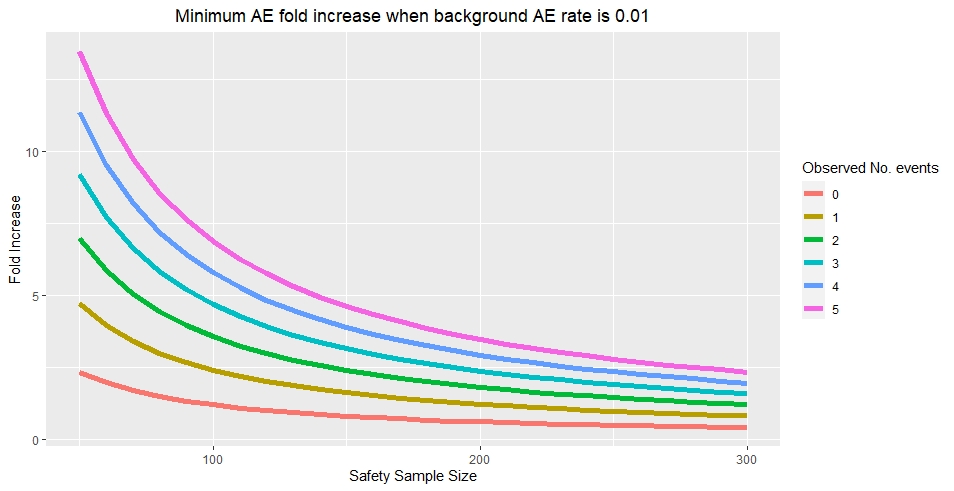}
        }%
        \subfigure[Probability threshold = 0.80]{%
            \includegraphics[width=3in]{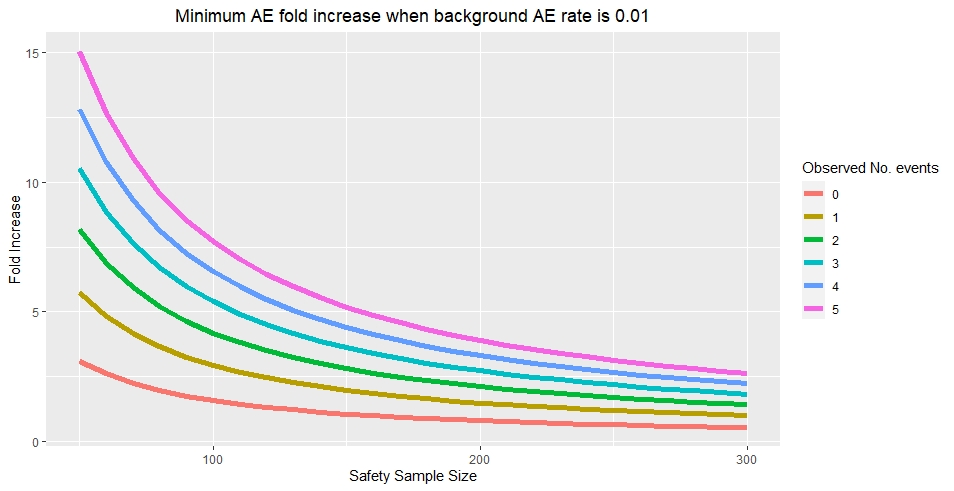}
        }\\
        \subfigure[Probability threshold = 0.90]{%
            \includegraphics[width=3in]{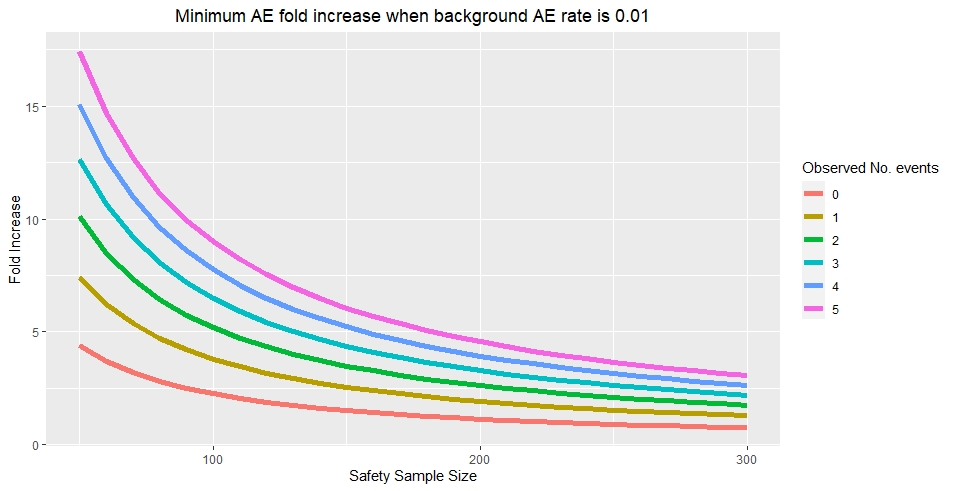}
        }%
        \subfigure[Probability threshold = 0.95]{%
            \includegraphics[width=3in]{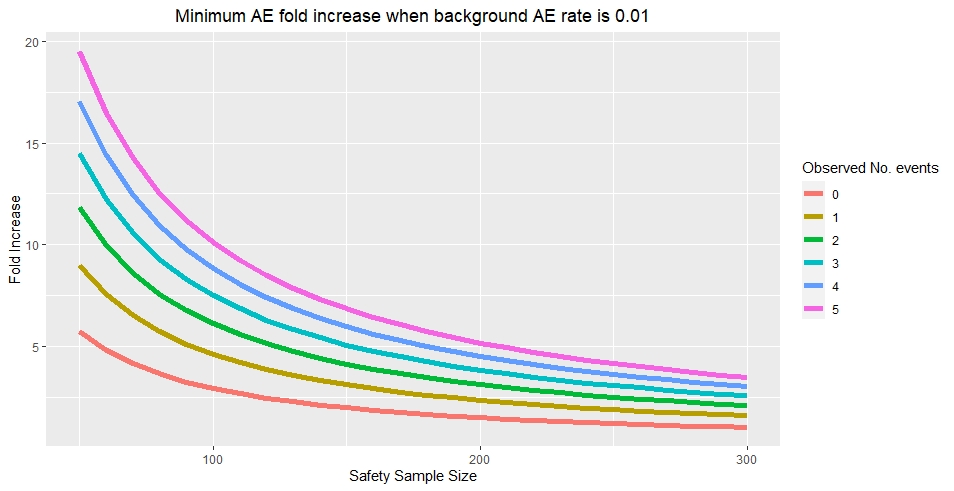}
        }\\
    \end{center}
    \caption{ Incidence rate cut off per observed incidence frequency as a function of sample size using different Probability thresholds.}%
   \label{fig:prob_cut_off}
\end{figure}

If $C$ were fixed at $0.80$, then we will be able to calculate $f$ given a reference background $\theta_1$ as a function of $(r_g, n_g)$. We will call this $f_C$ as the minimum fold based on probability threshold $C$. The minimum required number of patients with an AE before we can be certain that $p$ surpasses a specific threshold, or the maximum allowable number of patients with an AE after which we can no longer be certain that $p$ is below a certain threshold, given a certain confidence level (see Figure \ref{fig:prob_cut_off}). Figure \ref{fig:prob_cut_off} shows the fold increase in the specific AE rate when the background rate for that AE is fixed at $0.01$ at different confidence levels or probability thresholds $C$, i.e., $f_C$ as a function of $n_g$ for each possible observed $r_g$. For example, when the probability threshold is 80\%, if no event is observed for that specific AE, then we can rule out a doubling of the rate for that AE when the sample size is 150 participants. Of note, such a sample size is still relatively for many pediatric trials.  Once the probability threshold is raised, then the the doubling of the rate can no longer be ruled out. Furthermore, when the observed frequency of events increases, e.g., the more difficult it is to rule out a doubling of the incidence proportion even with a total sample size of 300 patients. In Figure \ref{fig:prob_cut_off}, it can be inferred that if the true incidence rate is small, the benefit gained in creasing sample size is not quite meaningful. On the other hand, if the true incidence rate is high, there is benefit in increasing sample size as the information (probability) gain for ascertaining $f$-fold increase is steep.  

\begin{figure}[ht!]
     \begin{center}
         \subfigure[True iTx inc. prop = 0.001]{%
            \includegraphics[width=3in]{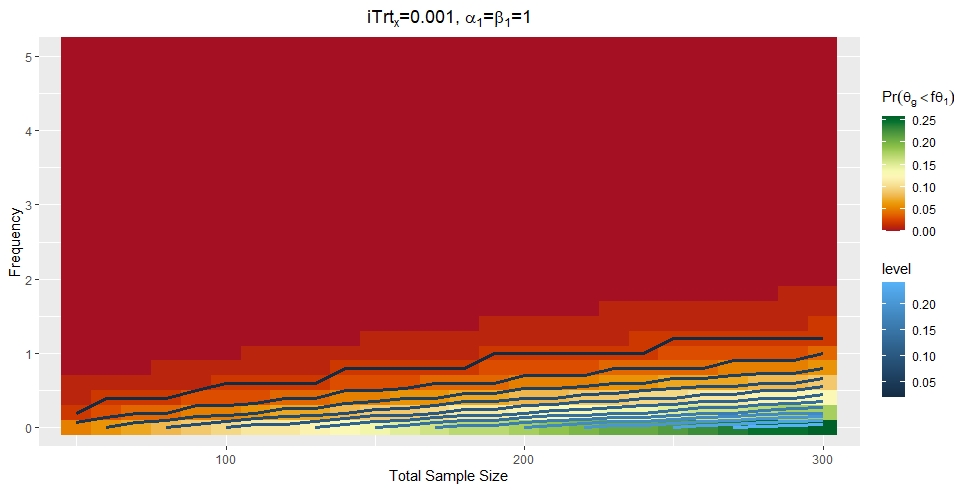}
        }%
        \subfigure[True iTx inc. prop = 0.0025]{%
            \includegraphics[width=3in]{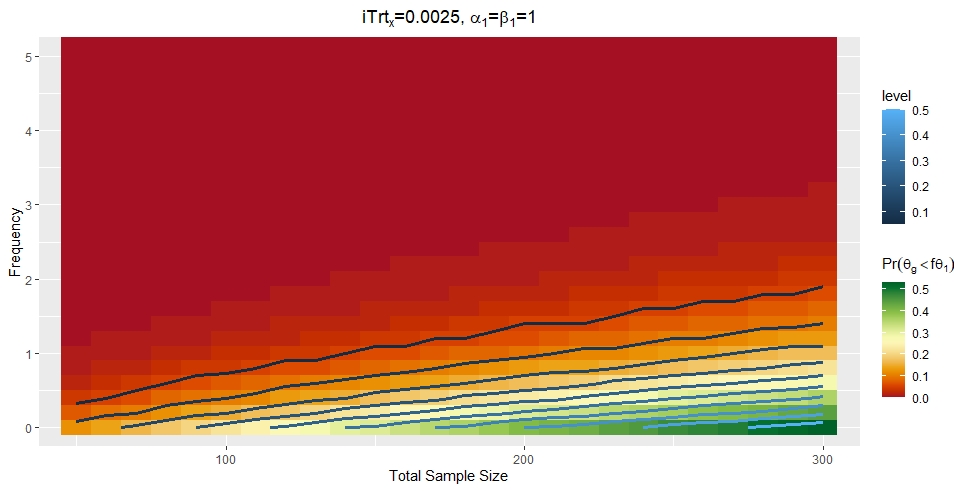}
        }\\
        \subfigure[True iTx inc. prop = 0.005]{%
            \includegraphics[width=3in]{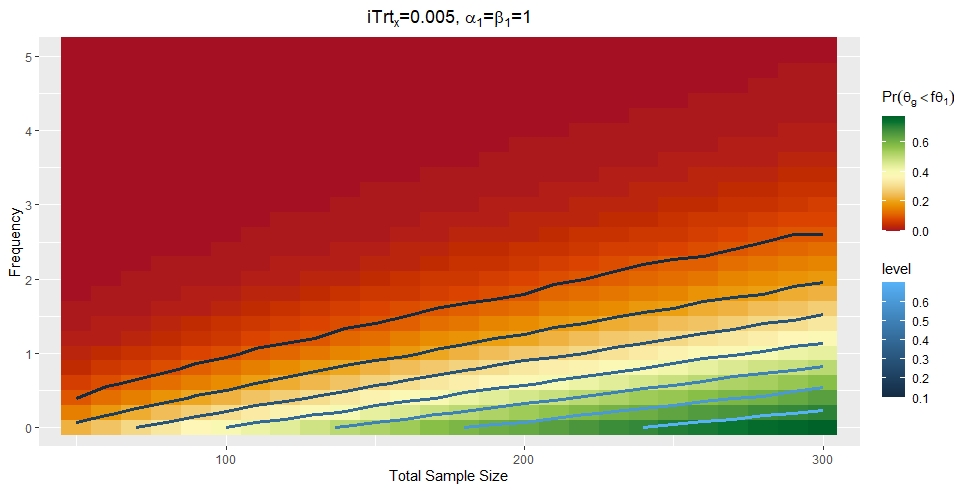}
        }%
        \subfigure[True iTx inc. prop = 0.01]{%
            \includegraphics[width=3in]{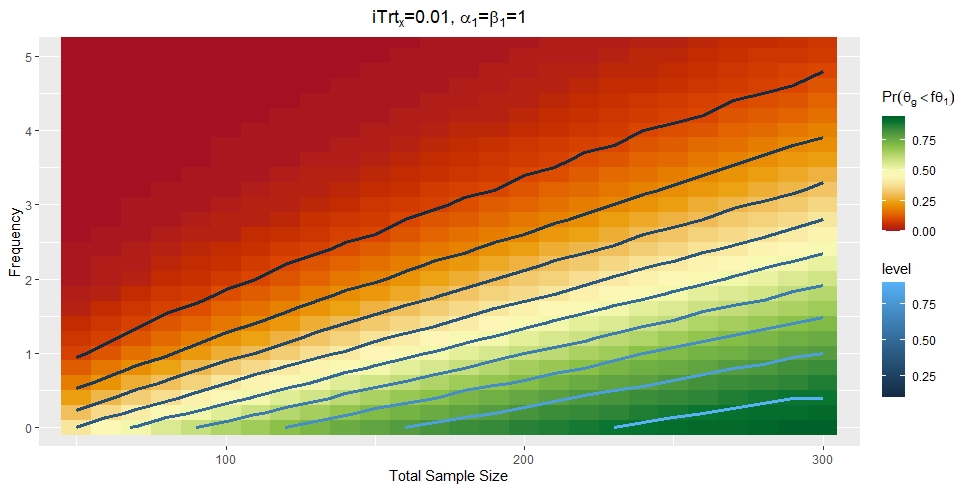}
        }\\
    \end{center}
    \caption{ Probability contour as a function of sample size and observed frequency of events.}%
   \label{fig:prob_contour}
\end{figure}

Ignoring $f$, Figure \ref{fig:prob_contour} shows the probability contours at different sample sizes and observed frequency of specific AEs, i.e., the figure shows the confidence level $C$ as a function of the plane created by $r_g$ and $n_g$. This then gives the likelihood of observing a particular frequency at a given sample size when the true incidence rate proportion is specified. At any of the sample sizes explored, and when the true incidence rate of the AE is small (less than 5/1000), the ability to observe at least 1 event is at most about 40\%. This maximum probability happens when the sample size is 300 patients. When the true incidence rate is 1/100, the probability of observing at least 1 event is ~60\%. Figure \ref{fig:prob_contour} shows that when the true incidence is small ($<1/100$) which is the upper bound for uncommon AEs the probability of observing at least one event in a 300 participant trial is low. Conversely, if one observes a supposedly uncommon event in the background population, that automatically gives a hint that the rate may be observed with a higher incidence proportion. In this example, we can recognize the importance of the Bayesian methodology for borrowing information from other populations or cohorts to increase the precision in the estimation of the incidence proportion of specific uncommon adverse events that are identified or are recognized as potential risks.   

\subsection{Precision for Developmental Safety Assessments}
The unique feature of assessing safety in children is in the consideration of developmental safety where there is relativity in pediatric growth data analysis. All age subgroups of the pediatric population are growing in size from the moment of birth until a certain time point during adolescence (or beyond adolescence for certain individuals) and each child grows at their own pace. Investigators should plan to evaluate the growth of the pediatric subjects in terms of anthropometric measurements. Usually, this includes the collection of data on sex, chronologic age (CA), bone age (BA), body weight, height, and head circumference and the extent of assessments may be influenced by the nature of the disease and the objective of a trial. In relation to these measurements, there is a need for outcomes that provide a sensitive measure of change since the trial cannot run for a long time and because growth happens slowly over time.  

Health care professionals sometimes convert height, weight or body mass index (BMI) measurements to standard deviation scores (SDS) and permits the understanding whether a measurement is normal by comparing it with the normal range of measurements for other children of the same age and sex.  \citep{cole1998british, inokuchi2011bmi}. A lower standard deviation score (in absolute value) means that the measurement is closer to the average or mean, while a high standard deviation score means that the value is further from the average or mean. A negative SDS indicates that the value is below the average or mean and a positive value means it is above the average or mean. The percentile number (or centile) shows how a child’s height compares to children of the same age and sex. 

For ascertaining the effect of the drug on growth measurements, the change from baseline SDS, i.e., $\mathbf{\Delta}SDS_{it_n}=SDS_{it_n}-SDS_{it_0}$, is determined over time, $t \in T=\left\{ t_0, \ldots, t_n\right\}$ and the expectation is that $E[\mathbf{\Delta}SDS_{it_n}]=0$. Typically, no formal hypothesis testing is done and conclusions are formed based on ocular inspection. Even if a hypothesis test is done, only the deviation at a landmark timepoint is considered $H_0: \mathbf{\Delta}SDS_{t_n} = 0$ vs. $H_1: \mathbf{\Delta}SDS_{t_n} < 0$, i.e., testing procedure generally starts with normality and in the process wants to rule out abnormality at $t_n$. However, not rejecting this hypothesis may just imply that the sample size is small. Hence, it may be more relevant to establish non-inferiority/consistency relative to placebo or context/benchmark for an development outcome, i.e., the realistic hypothesis is 
\begin{equation} H_{0,t_n}: \mathbf{\Delta}SDS_{t_n} \geq -\tau \quad vs.\quad H_{1,t_n}: \mathbf{\Delta}SDS_{t_n} < -\tau
\end{equation} where $\tau$ would correspond to a meaningful threshold. For example, suppose that the data in children pertaining to $\Delta SDS$ are $D= (x_1, \ldots, x_n)$ with $x_i \sim \mathcal{N}(\mu, \sigma^2)$. Let the prior $\pi(\mu, \sigma^2)= 1/\sigma^2$, then the posterior distribution for the mean change in SDS is the normal-inverse-$\chi^2$ $q(\mu, \sigma^2|D) = \mathcal{N}_{\chi^{-1}}(\bar{x}, n, n-1, s^2)$ where $\bar{x} = \sum_{i=1}^nx_i/n$ and $s^2 = \sum_{i=1}^n(x_i-\bar{x})^2/(n-1)$. Integrating out the variance $\sigma^2$ from the posterior distribution $q(\mu, \sigma^2|D)$, the assessment of the confidence level for a threshold $-\tau$ corresponds to the cumulative distribution of a non-central $t$ distribution with $n-1$ degrees of freedom, $t_{n-1}(\mu|\bar{x}, s^2/n)$. Because the non-central $t$ distribution is asymmetric and the mean is to the left of the median when $\bar{x}<0$, it is clear that the only way to rule out a shift of $\tau = 0.5$ with confidence level of at least 80\% when the size of the sample is 50 is if the observed mean change SDS is at least $\bar{x}\approx-0.04$. Furthermore, even if one increases the sample size to 100 the observed mean change SDS does not change appreciably to rule out a shift of 0.5 with confidence level of 0.80. In other words, from a statistical standpoint, it is challenging to rule out deleterious shifts even if we do not observe a shift in the mean. 

More realistically, however, the hypothesis test that corresponds to the meaningful clinical question is whether maximum of the average absolute change in SDS across time deviates from 0 as it is not known when the effect on development happens, i.e., if $\tilde{\Delta} SDS= \mathrm{max}_{t\in T} \Delta SDS_{t_n}$, we are interested in showing whether 
\begin{equation} 
H_{0}: \tilde{\Delta}SDS \geq -t\quad vs.\quad H_{1}: \tilde{\Delta}SDS < -t. 
\end{equation} It is known that for $x_1, \ldots, x_n$ that are independent and identically normally distributed with $N(\mu, \sigma^2)$, the probability density function is $f(x; \mu, \sigma)=n\sigma^{-1}\phi((x-\mu)/\sigma)\Phi((x-\mu)/\sigma)^{n-1}$. Suppose this assumption of independence is reasonable over time, the confidence level can be written similar to (\ref{eq: single_beta}) upon specification of the prior distributions for $\mu$ and $\sigma^2$ and calculation of the posterior distribution. This will be explored more in a subsequent research. In addition, it may be good to look at the increase in the variance and whether there is dose dependency in the disruption.  

Because it may be challenging to create a meaningful threshold for a deleterious shift in mean of SDS and because while the means may not change, the drug can cause shifts without increasing the mean or variances, i.e., the effect of the drug on the growth can be quite different for each individual. In this case, it may be informative to examine shifts of SDS over time in patients. Intervals at baseline may be defined into eight SD groups as described in \citet{tanaka2021changes} so that $z_{i0}=j$ if $\delta_{j-1}<y_{i0} <\delta_j$, $j=1, \ldots, 10$, $i=1, \ldots, n$, and $\{\delta_1, \delta_2, \ldots, \delta_9\} = \{ -\infty, -3.5, -2.5, -1.5, -0.5, 0.5, 1.5, 2.5, 3.5, \infty \}$. Subsequent evaluation of change in the height SDS from baseline to landmark times of assessment can be grouped into the following five SD groups, i.e., $d_{ik}=z_{ik}-z_{i0}=j$ if $\eta_{j-1}<y_{ik} <\eta_j$, $j=1, \ldots, 6$ and $i=1, \ldots, n$, and $\{\eta_1, \eta_2, \ldots, \eta_6\} = \{ -\infty, -1.5, -0.5, 0.5, 1.5, \infty \}$. From here, a Bayesian probit regression can be used for estimation and inference. 

Assessments for other measures, e.g., Tanner staging describing secondary sex characteristics for assessing maturational age, Bayley III Scales of Infant and Toddler Development to measure motor, cognitive, language, social-emotional, and adaptive behavior development in babies and young children for assessing normal neonatal and infant neurocognitive development \citep{choi2022} can be analyzed using through usual analytic strategy for dichotomous or polytomous endpoints. Nevertheless, a meaningful hypothesis or scientific question is necessary to facilitate the proper interpretation of the results. 

\section{Exploiting Archetypes of Pediatric Clinical Development for Informing Safety}

In terms of efficacy, the extent of pediatric clinical development is also influenced by whether extrapolation of efficacy can be supported or whether the ``line of reasoning'' has been established in the population intended for its use. Establishing the line of reasoning can be related to the current treatment landscape, referred henceforth as archetypes, for which the product belongs to at the time of pediatric study planning as proposed by \citet{gamalo2022extrapolation}. In particular, these archetypes are described as follows:  
\begin{itemize}
\item[A.] First-in-class or first-in-indication in adults:  In this archetype, there are no approved drugs in adults. However, there may be Phase 2 studies in adults establishing proof of direct benefit for the drugs in the class. There are no studies for the class of drugs in children.  
\item[B.] Established class where adult trials have been conducted: In this archetype, there are approved drugs in the same class in adults. There are no approved drugs in the class for children. 
\item[C.] Established class where adult and pediatric trials have been conducted: In this archetype, there are approved drugs in the same class for adults and children. 
\end{itemize}
Table \ref{table:1} gives some examples of drugs that possibly belong in these archetypes at the time their clinical development is planned or executed. For example, finerenone is an investigational, non-steroidal, selective mineralocorticoid receptor antagonist (MRA) that has been shown to reduce many of the harmful effects of mineralocorticoid receptors (MR) overactivation. This is a new mechanism of action for the treatment of type II diabetes and therefore belongs to Archetype A. In the migraine indication, when rimegepant, oral calcitonin gene-related peptide (CGRP) antagonist, started pediatric clinical development, another drug of the same mechanism of action has been approved in adults previously. Hence, it belongs to Archetype B. In both Archetype A and B, the line of reasoning in children for that mechanism of action has not yet been established. The only difference is that proof of direct benefit in Archetype B is broadly supported since it has also been shown effective in other drugs in the same class. For these archetypes, the requirement for demonstration of efficacy in a robust manner for children is needed and that the extent of safety database may also be large.  

Drugs being developed for children where only PK/PD or dose ranging study and a safety study is required most likely belongs to Archetype C as in the case of many anti-convulsants as treatment for partial onset seizures. To gain precision for the estimation of adverse reactions in compounds under this archetype can use of near-informative priors as the mechanism of action is predominantly responsible for many adverse reactions. For example, brivaracetam, a chemical analog of levetiracetam, was approved on August 27, 2021 for ages 1 month and older. In the development of brivaracetam, the extent of the safety database in children was from three single-arm open label studies: N01263 and N01266 with $n=346$ to cover the expanded age indication for the oral formulation while the review of study EP0065 ($n=50$) will examine all patients for coverage of the expanded indication for the intravenous formulation, 1 month to $<$ 16 years \citep{FDA_sNDA205836}. Levetiracetam, on the other hand, was approved for the same indication in children aged 4 to 16 on June 21, 2005 with a double-blind randomized placebo-control trial in $n=198$ \citep{FDA_sNDA21-035} and for ages 1 month to 4 years on December 16, 2011 forming a line of reasoning for extrapolation in this disease. The development in support of the approval for ages 1 month to 4 years includes 4 studies, including a double-blind randomized placebo-control trial in $n=116$, two open label long term safety studies $n=255$ for 1 year and $n=223$ for up to 7.5 years \citep{FDA_sNDA21035}. In fact, one can argue that for drugs in Archetype C, the extent of safety database is smaller unless a broader indication is sought after or that a certain unique signal is seen that may be of importance in children. Furthermore, the requirement for a double-blind randomized placebo-control trial is absent. 

\begin{table}[h!]
\centering
\begin{tabular}{c c c c} 
 \hline\hline
 Indication & Archetype A & Archetype B & Archetype C \\ [0.5ex] 
 \hline
 Migraine &  & rimegepant & lasmiditan \\ 
 Partial Onset Seizures & perampanel &  & brivaracetam \\ 
 Type II Diabetes & finerenone &  & lixisenatide  \\ 
 Atopic dermatitis & upadacitinib & lebrikizumab & \\
 RR-MS & evobrutinib & ocrelizumab & siponimod\\
 \hline
\end{tabular}
\caption{Archetypes of at time of development and example drugs}
\label{table:1}
\end{table}

For more broader indications (even if related) where no other drugs of similar mechanism have been studied or when a safety concern exists, the extent of clinical development may be larger to support both efficacy and safety in these groups of patients. For example, in the agreed Pediatric Investigational Plan for escarbazepine acetate, a voltage-gated sodium channel (VGSC) blocker chemically related to carbamazepine and oxcarbazepine, the extent of development in children included 8 double blind RCTs and open label RCTs in \citep{EMA_escar_PIP}. Similarly, developments tend to take very similar large extent, if the line of reasoning has not been fully established, e.g., conduct of adequately powered randomized controlled trials treatment and an open label extension in trials for the prevention of migraine headaches - a disease where the line of reasoning has not been established in children. For example, the extent of development for lasmiditan and rimegepant seem similar despite the fact that lasmiditan belongs to Archetype C. In this case, the development becomes similar to drugs falling in Archetype A which generally needs a PK + randomized controlled efficacy and/or safety study. There are cases for drugs in Archetype C where even if the line of reasoning has been established in children, and conceivably the development could be smaller, the actual extent of safety database still follows the traditional paradigm, e.g., ofatumumab for relapsing remitting multiple sclerosis. 

Different geographies have diverging views on being able to leverage existing safety from other relevant populations and indications. In the EU, the extrapolation guidance allows extrapolation of short-term safety data from adults to children \citep{european2012reflection}. In this case, Archetype C provides further opportunity in leveraging existing information for on-target effects. Further reduction in the size of safety and perhaps one can use registry all together. In the US FDA, on the other hand, there is an understanding that safety cannot be extrapolated as adverse effects can be different in children than from adults \citep{FDA_pediatric_rule}. Hence, even if there is a possibility to extrapolate, the extent of safety database will drive developments. Arguably, however, the adoption of the proposed landscape, e.g., Archetype C, implicitly allows for some extrapolation of safety from other populations, e.g. adults, and the realization that development, particularly may not be competitive once similar drugs are in the market. The use of these archetypes together with the concept of on-target effects could provide a strong argument for extrapolation of safety and in the use of quantitative measures to increase precision of estimates of incidence rates both cor common ADRs or for identified or potential risks that are less common as the use of informative priors can be further facilitated. 

\section{Discussion}

One objective of the FDA’s evaluation of safety in pediatric studies is to determine whether a product’s labeling needs to be revised to communicate to prescribers the benefit-risk information in children. In the reviews that have been done for investigational drugs for pediatric indications, some may conclude that the safety profile for children differs from that for adults in ways that, at a minimum, warranted discussion in the product’s labeling. In others, reviewers explicitly note whether the findings of treatment-related adverse events in children are reflected in the existing labeling (particularly for previously approved products) or whether some revisions are needed (see for e.g., \citep{field2012safety}). There are also times where the FDA reviewer does not compare pediatric safety findings to adult safety findings or make other appropriate comparisons (e.g., with findings for a control group or with safety findings in other pediatric studies of similar drugs for the same condition), while some reviews state only that no unexpected adverse events had been noted (see for e.g.). In context, such statements probably can be interpreted as suggesting that the safety profile was similar to that for adults if the product had been previously studied in adults \citep{boat2012safe}. Other reviews may indicate unique safety concerns on musculoskeletal, neurocognitive, and maturation either as indicated in relevant toxicology studies and/or outcomes in clinical studies. 

Almost all pediatric clinical trials are not designed to test specific hypotheses about safety nor to measure or identify adverse reactions with any pre-specified level of sensitivity. But often a requirement is imposed on the extent of the safety database without sufficient scientific rationale on the specific hypothesis to be addressed. Even with such a requirement, these databases may still not be adequate to answer objectives pertaining to differences in safety profiles and permit revision of labeling. This requirement may also wipe out the efficiency gained in demonstrating efficacy of the drug via efficient means. The use of on-target safety, where mechanistic explanation can supplement statistical uncertainty presents a good opportunity for a streamlined safety database that has an opportunity to answer important questions on safety without being too arbitrary as to extent of safety database. Furthermore, it also fits with the principle of right patients and right drug and is a concept that is consistent with extrapolation. Of note, the ICH E11 (R1) defines pediatric extrapolation (for purposes of this document, simply ``extrapolation'') as an approach to providing evidence in support of effective and safe use of drugs in the pediatric population when it can be assumed that the course of the disease and the expected response to a medicinal product would be sufficiently similar in the pediatric and reference population (e.g., adult, or other pediatric). On-target effects will be similar between adults and children; and hence, data can be extrapolated to optimize use of relevant information toward the pediatric population. Hence, this facilitates the use of Bayesian methods which provides a convenient approach for assessing probabilistic similarity and potential increase of precision needed in subsequent studies. Because posterior distributions contain all the relevant information about a quantity of interest, they provide an intuitive representation of a prediction and the associated tolerable uncertainty, enabling better decisions. Of note, even in the case for bizarre effects that are still target related, the use of Bayesian approach is still warranted.  

For a clear quantitative assessment of incidence rates of AE, it is important to have the appropriate hypothesis or scientific question in the assessment of safety so that conclusions made are sensible. To assess these hypotheses, we have laid out a unified expression for the assessment of safety incidence rates in relation to confidence levels through posterior distributions of certain quantities of interest. This expression provides a relationship between the confidence level for ruling out certain placebo adjusted AE rate, a multiple of the placebo adjusted AE rate, sample size, observed event rates, and priors. By using on-target effects, it is possible to leverage information from related sources. Only in cases where we can use available related information can meaningful conclusions of safety happen. Otherwise, there will be a non-negligible uncertainty in these adverse reactions.  

Archetypes of pediatric clinical development provide useful information to support extrapolation both from the efficacy and safety perspective and can be used to streamline questions and facilitate efficient use of data in an ecosystem. In established classes, i.e., Archetype B and C, the line of reasoning that is created gives further support for the ability to leverage data through Bayesian statistics or other meta-analytic tools need to be explored to speed development and ensure feasibility of development where there is competing availability of drugs in the market. Alternative strategies in the post-market utilizing multi-sponsor master protocols can leverage a common placebo control and/or gain efficiency through leveraging of data from similar compounds.  

There are other questions that will need to be addressed in the future in relation to streamlined and quantitative assessments of safety. In many cases, detecting rare adverse drug reactions with latent onset may not be ruled out because trials are too short in duration to capture ADRs associated with long-term use. Unknown effects not seen in pivotal trials may be better characterized in post-market settings with appropriate mechanisms for rapid reporting. Furthermore, comparative data in pediatrics may not be available for proper characterization from background rates in the population due to the feasibility of exposing children to no treatment for an extended period. Of note, comparative safety data is of importance when the background rate of important ADRs is high and well-established treatment already exists with effect on a clinically meaningful outcome. It is hoped that all these questions can be catalogued, and proper statistical methods identified so that assessments of uncertainty can be clearly determined.

\bibliographystyle{unsrtnat}
\bibliography{ref_adp}
\end{document}